\documentclass{rnaastex}

\usepackage{amsmath}
\usepackage{url}
\usepackage{xspace}

\newcommand{\sdssz}{SDSS \textsl{z$^\prime$}\xspace}

\begin{document}

\title{Photometric Analysis \& Transit Times of TRAPPIST-1 \MakeLowercase{b}, \MakeLowercase{c}}

\author[0000-0003-2528-3409]{Brett M. Morris}
\affiliation{Astronomy Department, University of Washington, Seattle, WA 98195, USA}

\author{Eric Agol}
\affiliation{Astronomy Department, University of Washington, Seattle, WA 98195, USA}

\author{Suzanne L. Hawley}
\affiliation{Astronomy Department, University of Washington, Seattle, WA 98195, USA}

\email{bmmorris@uw.edu}

\accepted{RNAAS, January 10, 2018}


\maketitle 

TRAPPIST-1 hosts seven Earth-sized planets transiting an M8 star \citep{Gillon2016, Gillon2017}. We observed mid-transit times of each of the inner two planets with the Astrophysical Research Consortium (ARC) 3.5 m Telescope at Apache Point Observatory (APO) to help constrain the planet masses with transit timing variations \citep{Agol2005, Holman2005}, and we outline a procedure for analyzing transit observations with APO.

We observed one transit each of TRAPPIST-1 b and c with the ARCTIC imager using 4$\times$4 binning and 10 second exposures in \sdssz \citep{Huehnerhoff2016}. Since the star is dim ($R = 16.6$) and the transit depths are small ($\Delta F/F \sim 1\%)$, we develop a technique for removing background fringing, and for computing the transit light curve. 

We collected ten night sky flats, rotating and slewing the telescope between exposures, to correct for background fringing in the \sdssz-band \citep{Howell2006}. Typically, flat fields are created by taking the median of a number of exposures of a constant, evenly illuminated field. The sky emission that produces the fringe pattern is variable and non-monotonic, so we developed a technique for flats.

In the series of $i=1,...,N$ flats, the flux of the $j^\textsuperscript{th}$ pixel, $p_{j,i}$ changes significantly between frames. Occasionally a star falls on a pixel, making one or two fluxes orders of magnitude brighter than the others. The median of all pixels in the $i^\textsuperscript{th}$ frame, $m_i$, produces a light curve, which tracks the variations in flux from the sky emission. We regress $p_{j,i}$ against $m_i$  to match each individual pixel's light curve, masking $3\sigma$ outliers when a star landed within a pixel. We normalize the matrix of linear regression coefficients by its median to create the flat field. 

We correct for the local background measured in circular annuli centered on the stellar centroids using annuli with radii 10 and 20$\times$ larger than each source aperture radius. We normalize the TRAPPIST-1 light curve by a mean comparison star, which is computed from a linear combination of the following regressors: the fluxes of each comparison star, the target centroid pixel $x$ and $y$ coordinates, median sky background, air humidity, air pressure, and airmass. We regress the light curve of TRAPPIST-1 against these time series, and normalize the light curve by the combination of regressors that minimizes the out-of-transit scatter in the light curve. To avoid overfitting, we used the principle component analysis (PCA) cross-validation technique of \citet{luger2016}, which reduces the large number of available regressors to the smallest number of significant principle components necessary to detrend the light curve. We train the regression on a fraction of the out-of-transit observations while excluding a number of test fluxes. We choose the number principle components and aperture radius that produce a mean comparison star light curve with minimal test flux scatter. 

\begin{figure*}
\begin{center}
\includegraphics[scale=0.6]{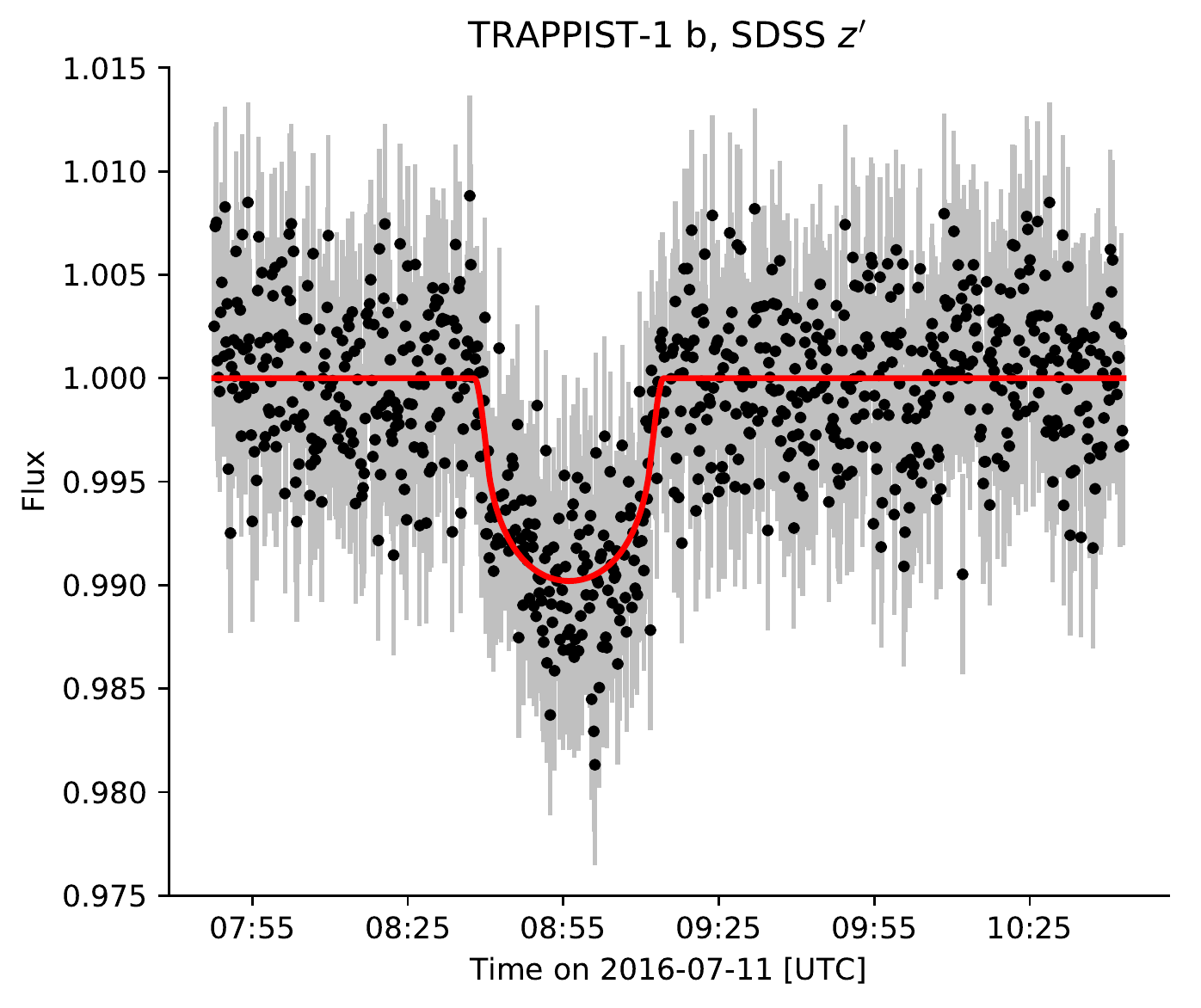}
\includegraphics[scale=0.6]{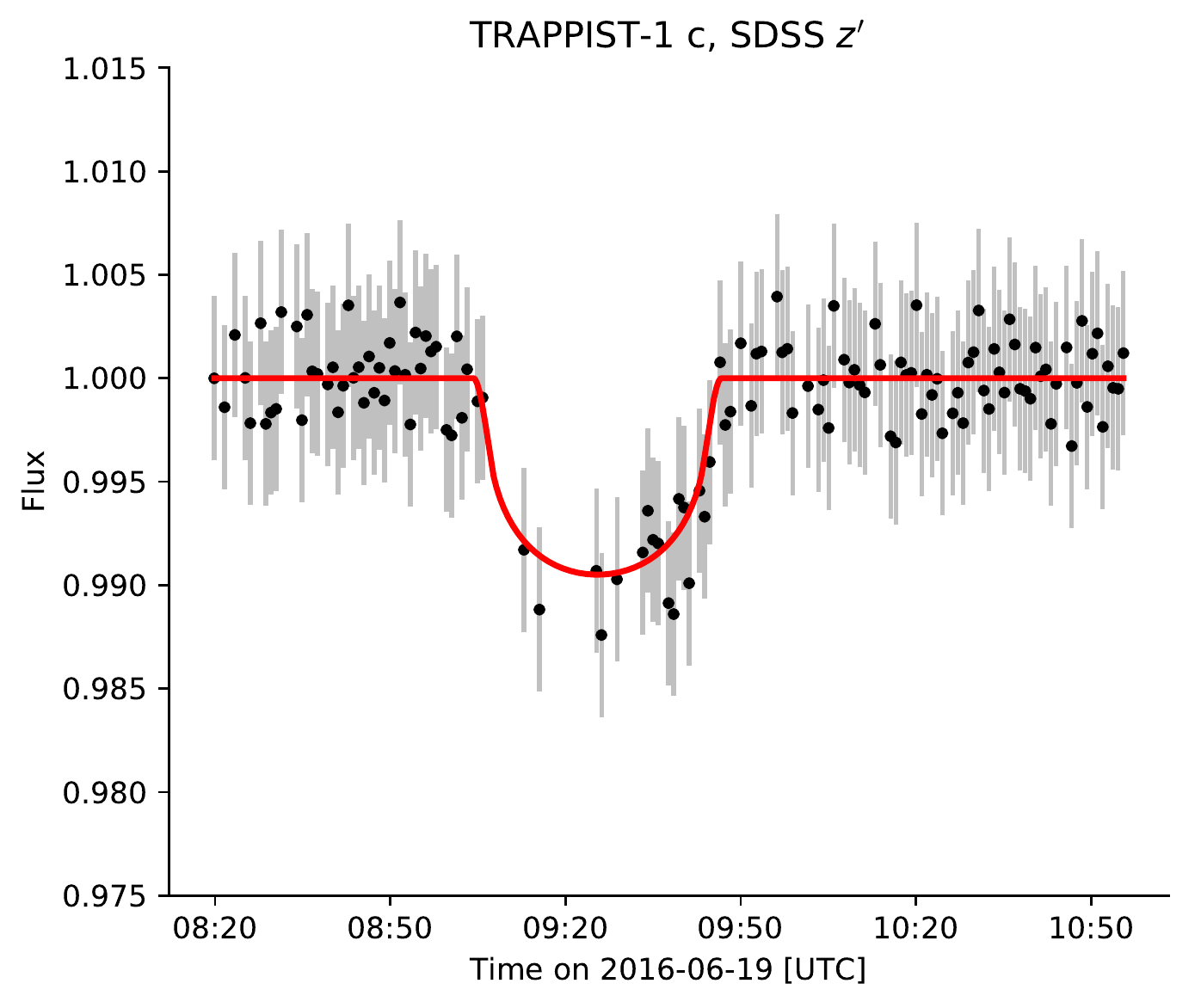}
\end{center}
\caption{Transits of TRAPPIST-1.\label{fig:transits}}
\end{figure*}

We fit the light curve in Figure~\ref{fig:transits} for the depth and mid-transit time, and fix other parameters to the values of \citet{Gillon2016}, with quadratic limb-darkening \citep{mandel2002}. We compute $N$ posterior samples with Markov Chain Monte Carlo \citep{Foreman-Mackey2013}, and model correlated noise with a Matern 3/2 gaussian process \citep{Ambikasaran2014}. Chains are ``converged'' when $N > 150 \tau_{int}$, where $\tau_{int}$ is the integrated autocorrelation length. Analysis software and posterior samples are available online \citep{software}.

The transit times of TRAPPIST-1 b and c are $\mathrm{BJD}_{\mathrm{TDB}} = 2457580.87634^{+0.00034}_{-0.00034}$ and
$2457558.89477^{+0.00080}_{-0.00085}$, respectively, which will help constrain the planet masses.

\acknowledgments

We gratefully acknowledge help from Rodrigo Luger and Dan Foreman-Mackey. Based on observations obtained with the APO 3.5-meter telescope, which is owned and operated by ARC.

\software{\texttt{trappist1\_arctic\_2016} \citep{software}, \texttt{astropy} \citep{Astropy2013}, \texttt{photutils} \citep{Bradley2016}, \texttt{george} \citep{Foreman-Mackey2015}, \texttt{emcee} \citep{Foreman-Mackey2013}, \texttt{astroplan} \citep{astroplan}}

\facility{APO/ARC 3.5m}


\begin{thebibliography}{}
\expandafter\ifx\csname natexlab\endcsname\relax\def\natexlab#1{#1}\fi
\providecommand{\url}[1]{\href{#1}{#1}}

\bibitem[{{Agol} {et~al.}(2005){Agol}, {Steffen}, {Sari}, \&
  {Clarkson}}]{Agol2005}
{Agol}, E., {Steffen}, J., {Sari}, R., \& {Clarkson}, W. 2005, \mnras, 359, 567

\bibitem[{{Ambikasaran} {et~al.}(2014){Ambikasaran}, {Foreman-Mackey},
  {Greengard}, {Hogg}, \& {O'Neil}}]{Ambikasaran2014}
{Ambikasaran}, S., {Foreman-Mackey}, D., {Greengard}, L., {Hogg}, D.~W., \&
  {O'Neil}, M. 2014, ArXiv e-prints, arXiv:1403.6015

\bibitem[{{Astropy Collaboration} {et~al.}(2013){Astropy Collaboration},
  {Robitaille}, {Tollerud}, {Greenfield}, {Droettboom}, {Bray}, {Aldcroft},
  {Davis}, {Ginsburg}, {Price-Whelan}, {Kerzendorf}, {Conley}, {Crighton},
  {Barbary}, {Muna}, {Ferguson}, {Grollier}, {Parikh}, {Nair}, {Unther},
  {Deil}, {Woillez}, {Conseil}, {Kramer}, {Turner}, {Singer}, {Fox}, {Weaver},
  {Zabalza}, {Edwards}, {Azalee Bostroem}, {Burke}, {Casey}, {Crawford},
  {Dencheva}, {Ely}, {Jenness}, {Labrie}, {Lian Lim}, {Pierfederici},
  {Pontzen}, {Ptak}, {Refsdal}, {Servillat}, \& {Streicher}}]{Astropy2013}
{Astropy Collaboration}, {Robitaille}, T.~P., {Tollerud}, E.~J., {et~al.} 2013,
  \aap, 558, A33

\bibitem[{{Bradley} {et~al.}(2016){Bradley}, {Sipocz}, {Robitaille},
  {Tollerud}, {Deil}, {Vin{\'{\i}}cius}, {Barbary}, {G{\"u}nther}, {Bostroem},
  {Droettboom}, {Bray}, {Bratholm}, {Pickering}, {Craig}, {Pascual}, {Greco},
  {Donath}, {Kerzendorf}, {Littlefair}, {Barentsen}, {D'Eugenio}, \&
  {Weaver}}]{Bradley2016}
{Bradley}, L., {Sipocz}, B., {Robitaille}, T., {et~al.} 2016, {Photutils:
  Photometry tools}, Astrophysics Source Code Library, , , ascl:1609.011

\bibitem[{{Foreman-Mackey}(2015)}]{Foreman-Mackey2015}
{Foreman-Mackey}, D. 2015, {George: Gaussian Process regression}, Astrophysics
  Source Code Library, , , ascl:1511.015

\bibitem[{{Foreman-Mackey} {et~al.}(2013){Foreman-Mackey}, {Hogg}, {Lang}, \&
  {Goodman}}]{Foreman-Mackey2013}
{Foreman-Mackey}, D., {Hogg}, D.~W., {Lang}, D., \& {Goodman}, J. 2013, \pasp,
  125, 306

\bibitem[{{Gillon} {et~al.}(2016){Gillon}, {Jehin}, {Lederer}, {Delrez}, {de
  Wit}, {Burdanov}, {Van Grootel}, {Burgasser}, {Triaud}, {Opitom}, {Demory},
  {Sahu}, {Bardalez Gagliuffi}, {Magain}, \& {Queloz}}]{Gillon2016}
{Gillon}, M., {Jehin}, E., {Lederer}, S.~M., {et~al.} 2016, \nat, 533, 221

\bibitem[{{Gillon} {et~al.}(2017){Gillon}, {Triaud}, {Demory}, {Jehin}, {Agol},
  {Deck}, {Lederer}, {de Wit}, {Burdanov}, {Ingalls}, {Bolmont}, {Leconte},
  {Raymond}, {Selsis}, {Turbet}, {Barkaoui}, {Burgasser}, {Burleigh}, {Carey},
  {Chaushev}, {Copperwheat}, {Delrez}, {Fernandes}, {Holdsworth}, {Kotze}, {Van
  Grootel}, {Almleaky}, {Benkhaldoun}, {Magain}, \& {Queloz}}]{Gillon2017}
{Gillon}, M., {Triaud}, A.~H.~M.~J., {Demory}, B.-O., {et~al.} 2017, \nat, 542,
  456

\bibitem[{{Holman} \& {Murray}(2005)}]{Holman2005}
{Holman}, M.~J., \& {Murray}, N.~W. 2005, Science, 307, 1288

\bibitem[{{Howell}(2006)}]{Howell2006}
{Howell}, S.~B. 2006, {Handbook of CCD Astronomy}, ed. R.~{Ellis}, J.~{Huchra},
  S.~{Kahn}, G.~{Rieke}, \& P.~B. {Stetson}

\bibitem[{{Huehnerhoff} {et~al.}(2016){Huehnerhoff}, {Ketzeback}, {Bradley},
  {Dembicky}, {Doughty}, {Hawley}, {Johnson}, {Klaene}, {Leon}, {McMillan},
  {Owen}, {Sayres}, {Sheen}, \& {Shugart}}]{Huehnerhoff2016}
{Huehnerhoff}, J., {Ketzeback}, W., {Bradley}, A., {et~al.} 2016, in \procspie,
  Vol. 9908, Ground-based and Airborne Instrumentation for Astronomy VI, 99085H

\bibitem[{{Luger} {et~al.}(2016){Luger}, {Agol}, {Kruse}, {Barnes}, {Becker},
  {Foreman-Mackey}, \& {Deming}}]{luger2016}
{Luger}, R., {Agol}, E., {Kruse}, E., {et~al.} 2016, \aj, 152, 100

\bibitem[{{Mandel} \& {Agol}(2002)}]{mandel2002}
{Mandel}, K., \& {Agol}, E. 2002, \apjl, 580, L171

\bibitem[{Morris(2017)}]{software}
Morris, B.~M. 2017, doi:10.5281/zenodo.1064302

\bibitem[{{Morris} {et~al.}(2017){Morris}, {Tollerud}, {Sipocz}, {Deil},
  {Douglas}, {Berlanga Medina}, {Vyhmeister}, {Smith}, {Littlefair},
  {Price-Whelan}, {Gee}, \& {Jeschke}}]{astroplan}
{Morris}, B.~M., {Tollerud}, E., {Sipocz}, B., {et~al.} 2017, ArXiv e-prints,
  arXiv:1712.09631

\end{thebibliography}
\end{document}